\documentclass[11pt,twocolumn]{article}

\usepackage[margin=1in]{geometry}
\usepackage[T1]{fontenc}
\usepackage[utf8]{inputenc}
\usepackage{times}
\usepackage{url}
\usepackage{amsmath,amssymb}
\usepackage{booktabs}
\usepackage{graphicx}
\usepackage{enumitem}
\usepackage{longtable}
\usepackage{microtype}

\usepackage{hyperref}
\hypersetup{
  colorlinks=true,
  linkcolor=black,
  citecolor=black,
  urlcolor=blue
}

\title{Auto-Tuning Safety Guardrails for Black-Box Large Language Models}

\author{
  Perry Abdulkadir\thanks{Work completed as part of the M.S.\ in Artificial Intelligence at the University of St.\ Thomas, using publicly available models and datasets. The views expressed and any errors are the author's own.} \\
  University of St.\ Thomas \\
  \texttt{abdu9698@stthomas.edu} \\
  \texttt{perry.abdulkadir@gmail.com}
}

\date{December 10, 2025}

\begin{document}

\maketitle

\begin{abstract}
Large language models (LLMs) are increasingly deployed behind safety guardrails such as system prompts and content filters, especially in settings where product teams cannot modify model weights. In practice these guardrails are typically hand-tuned, brittle, and difficult to reproduce. This paper studies a simple but practical alternative: treat safety guardrail design itself as a hyperparameter optimization problem over a frozen base model. Concretely, I wrap Mistral-7B-Instruct with modular jailbreak and malware system prompts plus a ModernBERT-based harmfulness classifier, then evaluate candidate configurations on three public benchmarks covering malware generation, classic jailbreak prompts, and benign user queries. Each configuration is scored using malware and jailbreak attack success rate, benign harmful-response rate, and end-to-end latency. A 48-point grid search over prompt combinations and filter modes establishes a baseline. I then run a black-box Optuna study over the same space and show that it reliably rediscovers the best grid configurations while requiring an order of magnitude fewer evaluations and roughly 8$\times$ less wall-clock time. The results suggest that viewing safety guardrails as tunable hyperparameters is a feasible way to harden black-box LLM deployments under compute and time constraints.
\end{abstract}

\section{Introduction}

Large language models (LLMs) are now embedded in productivity tools, coding assistants, educational applications, and many other high-impact products. In many real deployments, teams do not fine-tune the underlying model weights: the base LLM is provided as a managed service, or internal governance and infrastructure constraints make weight-level changes infeasible. Instead, safety is implemented through deployment-time guardrails such as system prompts, safety policies, and content filters wrapped around a frozen base model.

These guardrails matter: simple changes to system prompts, refusal templates, or classifier thresholds can significantly alter an application's vulnerability to jailbreaks and other misuse. However, they are typically tuned informally by a small set of practitioners, making them brittle, non-replicable, and hard to reason about.

This work investigates a simple question:

\medskip
\noindent
\textbf{Research question.} \emph{Given a frozen conversational LLM, can we automatically search over a discrete space of safety guardrail configurations to reduce safety failures while maintaining helpfulness and reasonable latency?}
\medskip

I present a small proof-of-concept system that treats guardrail design---specifically, combinations of safety-oriented system prompts and content-filter modes---as a hyperparameter optimization problem around a fixed base model. Rather than hand-tune these choices or exhaustively enumerate all combinations, I use standard black-box hyperparameter optimization to search for high-performing configurations.

Concretely, I:

\begin{itemize}[leftmargin=*]
  \item Wrap Mistral-7B-Instruct-v0.2\footnote{Model card for mistralai/Mistral-7B-Instruct-v0.2 on Hugging Face.} with modular jailbreak and malware system prompts plus a ModernBERT-based harmfulness classifier.\footnote{Model card for Jazhyc/modernbert-wildguardmix-classifier and AI2 WildGuardMix dataset card on Hugging Face.}
  \item Evaluate each configuration on three public datasets covering malware generation, classic jailbreak prompts, and benign behaviors.
  \item Define four metrics: malware attack success rate (ASR), jailbreak ASR, benign harmful-response rate (a proxy for over-refusal/hallucination), and end-to-end latency.
  \item Run a full 48-point grid search to establish a baseline, then apply Optuna\footnote{Optuna: A hyperparameter optimization framework documentation.} to search the same space more efficiently.
\end{itemize}

The experiments are small in scale but representative of realistic constraints: limited time and compute, no access to model gradients, and a limited guardrail configuration space. The main findings are:

\begin{enumerate}[leftmargin=*]
  \item Without any guardrails, the base model is highly vulnerable, especially to jailbreak prompts.
  \item Adding a simple classifier-based content filter meaningfully reduces attack success at modest latency cost.
  \item Combining system prompts with filtering yields better benign performance than filtering alone.
  \item Off-the-shelf hyperparameter optimization (Optuna) recovers high-performing guardrail configurations substantially faster than naive grid search.
\end{enumerate}

Although exploratory, these results support the view that safety guardrails can be treated as first-class tunable objects. The goal is not to propose a novel algorithm, but to demonstrate that standard, well-understood tools from hyperparameter optimization can be repurposed to help harden black-box LLM deployments in practice.

\section{Problem Setting}

I consider a standard chat-style LLM interface where a user issues a prompt $u$, the system wraps it with a system prompt $s$ and possibly other control tokens, and a frozen base model $f_\theta$ produces a response $r$:
\[
r = f_\theta(s, u).
\]

Safety guardrails are implemented via:

\begin{enumerate}[leftmargin=*]
  \item \textbf{System prompts:} additional natural-language instructions reminding the model to refuse unsafe requests (e.g., jailbreak or malware attempts).
  \item \textbf{Content filters:} a harmfulness classifier $g_\phi$ that scores $(u, r)$ and either passes through or overrides $r$ with a refusal if the predicted risk exceeds a threshold.
\end{enumerate}

In many product organizations, practitioners hand-tune $s$ and the filtering policy. Instead, I define a discrete configuration space $\mathcal{C}$ of possible guardrail choices and seek a configuration $c^\star \in \mathcal{C}$ that reduces safety failures while preserving benign helpfulness and keeping latency acceptable.

Formally, given evaluation datasets of \emph{malware prompts}, \emph{jailbreak prompts}, and \emph{benign prompts}, I define a vector-valued objective for configuration $c$:
\[
J(c) = \big(\text{ASR}_{\text{mal}}(c),\ \text{ASR}_{\text{jb}}(c),\ \text{Harm}_{\text{ben}}(c),\ \text{Latency}(c)\big),
\]
and aim to approximate Pareto-optimal trade-offs or minimize a scalarized objective constructed from these components. Because $f_\theta$ and $g_\phi$ are used as black boxes, I use black-box hyperparameter optimization to search over $\mathcal{C}$.

\section{Method}

\subsection{Base Model and Generation}

The base LLM is Mistral-7B-Instruct-v0.2, loaded via HuggingFace Transformers on an A100 GPU. I use the model in standard causal language modeling mode with a simple instruction-style prompt template:
\begin{verbatim}
<s> [INST] {system_prompt}
User: {user_prompt} [/INST]
\end{verbatim}

Generation hyperparameters are:
\begin{itemize}[leftmargin=*]
  \item \texttt{max\_new\_tokens} = 256
  \item \texttt{temperature} = 0.3, \texttt{top\_p} = 0.9 for general chat
  \item Slightly lower temperature (0.2) for pure code prompts
\end{itemize}

Latency is measured as wall-clock time per generation (including CUDA synchronization) for each prompt.

\subsection{Harmfulness Classifier and Filter Modes}

To assess and optionally filter responses, I use the ModernBERT-based \texttt{modernbert-wildguardmix-classifier}, a binary harmfulness classifier trained on safety-related text. A helper function
\texttt{classify\_exchange(prompt, response)} returns a probability $p_\text{harm}$ that the exchange is harmful.

On top of this classifier, I implement three content-filter modes:

\begin{itemize}[leftmargin=*]
  \item \textbf{none}: always return the raw LLM output.
  \item \textbf{mild}: if $p_\text{harm} \ge 0.5$, replace the output with a canned refusal.
  \item \textbf{strict}: if $p_\text{harm} \ge 0.8$, replace the output.
\end{itemize}

In all cases, the classifier is also used to compute safety metrics, as described below.

\subsection{Modular System Prompts}

On top of a brief ``helpful assistant'' base system prompt, I define four modular safety snippets, each of which can be toggled on or off:

\begin{itemize}[leftmargin=*]
  \item \textbf{JB1}: instructions not to follow any requests that attempt to bypass or ``jailbreak'' safety rules.
  \item \textbf{JB2}: explicit guidance not to comply with role-play or ``pretend'' instructions that simulate unsafe systems.
  \item \textbf{MW1}: a high-level prohibition on generating malware, exploit code, or step-by-step instructions for harmful activities.
  \item \textbf{MW2}: more concrete refusal patterns for harmful coding assistance (e.g., refusing to write payloads or exploits).
\end{itemize}

A guardrail configuration $c$ thus consists of binary flags for these four snippets plus a choice of content-filter mode.

\subsection{Datasets}

I use three existing datasets to stress different aspects of safety:

\paragraph{Malware prompts (RMCBench).}
A subset of 50 prompts from RMCBench,\footnote{GitHub repository for RMCBench (qing-yuan233, 2025) and Chen et al. (2024).} which contains adversarial prompts designed to elicit malicious code in both text-to-code and code-to-code settings. The key metric here is malware ASR.

\paragraph{Jailbreak prompts (ChatGPT-Jailbreak-Prompts).}
A subset of 50 classic jailbreak attempts from \texttt{rubend18/ChatGPT-Jailbreak-Prompts},\footnote{CyberAlbSecOP, ``Awesome GPT Super Prompting'' jailbreak collection on GitHub, and Jaramillo (2022).} including ``Developer Mode'' and other well-known system-prompt attacks. The metric is jailbreak ASR.

\paragraph{Benign prompts (JBB-Behaviors).}
Benign examples from the JBB-Behaviors split of JailbreakBench,\footnote{JailbreakBench (2025) and Chao et al. (2024).} which pairs unsafe behaviors with safe, everyday behaviors. I sample 50 benign prompts and measure how often the system produces outputs flagged as harmful.

All datasets are in English and drawn from public HuggingFace repositories.

\subsection{Metrics}

For a given configuration $c$, the evaluation harness computes:

\begin{itemize}[leftmargin=*]
  \item \textbf{Malware ASR}: fraction of malware prompts whose final responses are classified as harmful by the classifier.
  \item \textbf{Jailbreak ASR}: fraction of jailbreak prompts with harmful responses.
  \item \textbf{Benign harmful-response rate}: fraction of benign prompts whose responses are classified as harmful (proxy for over-refusal or hallucinated risk).
  \item \textbf{Latency}: average wall-clock time per prompt, including generation and filtering.
\end{itemize}

The bare configuration uses no safety snippets and \texttt{filter-none}, serving as a baseline for vulnerability and latency.

\subsection{Search Procedures}

\paragraph{Grid search.}

The grid search enumerates all 48 configurations obtained from the $2^4$ combinations of the four system-prompt snippets (JB1, JB2, MW1, MW2) crossed with the three filter modes (\texttt{none}, \texttt{mild}, \texttt{strict}). For each configuration, the system evaluates all three datasets and records the metrics defined above.

\paragraph{Optuna search.}

To explore whether black-box hyperparameter optimization can find good guardrail configurations more efficiently, I define an Optuna search space with:

\begin{itemize}[leftmargin=*]
  \item Binary variables for JB1, JB2, MW1, MW2.
  \item A categorical variable for filter mode in \{\texttt{none}, \texttt{mild}, \texttt{strict}\}.
\end{itemize}

The objective for Optuna is a scalarized score that combines the four metrics:
\begin{equation*}
\text{score}(c) =
\begin{aligned}[t]
  &0.4\,\text{ASR}_{\text{mal}} + 0.4\,\text{ASR}_{\text{jb}} \\
  &+ 0.1\,\text{Harm}_{\text{ben}} + 0.1\,\text{Latency}
\end{aligned}
\end{equation*}
where all terms are normalized. Lower values are better. For efficiency, each trial initially uses only 10 prompts per dataset; the top 5 configurations found by Optuna are then re-evaluated on the full 50-prompt sets.

\section{Experiments}

\subsection{Setup}

All experiments are run in Google Colab on an A100 GPU. For the grid search, each of the 48 configurations is evaluated on all 50 prompts from each dataset. For Optuna, I run 24 trials in the fast loop (10 examples per dataset) and then re-score the best 5 configurations on the full datasets.

\subsection{Baseline Vulnerability}

The bare configuration (no safety snippets, \texttt{filter-none}) illustrates the need for guardrails. Malware ASR is roughly 0.48 on the sampled RMCBench prompts, and jailbreak ASR is close to 0.98 on the jailbreak prompts: almost every attack elicits a harmful response. Benign harmful-response rate is also high (around 0.42), suggesting that the classifier tends to flag many raw outputs as problematic.

\subsection{Effect of Content Filtering}

Adding classifier-based filtering on top of the bare model already helps. Moving from \texttt{filter-none} to \texttt{filter-strict} reduces malware ASR by about 10 percentage points (0.48 $\rightarrow$ 0.38) at the cost of extra latency from classifier calls. Jailbreak ASR falls modestly, while benign harmful-response rate also changes, reflecting the classifier's impact on over-blocking.

\subsection{System Prompts + Filtering}

Configurations that combine safety-oriented system prompts with filtering tend to yield the best benign behavior. For example, turning on both jailbreak and malware snippets (JB1, JB2, MW1, MW2) and using mild filtering achieves benign harmful-response rates as low as roughly 0.22 in the grid search, while reducing malware ASR compared to the bare baseline. The main challenge is that jailbreak ASR remains high across many configurations, indicating that simple prompt reminders and a general-purpose harmfulness classifier are not enough to fully harden the model against prompt-injection attacks.

\subsection{Hyperparameter Optimization Results}

The Optuna study, though small, demonstrates that standard hyperparameter optimization can quickly discover good guardrail configurations:

\begin{itemize}[leftmargin=*]
  \item With only 24 trials evaluated on 10 prompts per dataset, Optuna converges on configurations that mirror the best-performing ones from the full grid search.
  \item The best Optuna trial favors a combination of specific jailbreak and malware prompts with mild filtering; re-evaluated on the full datasets, its safety and latency metrics closely match or slightly improve on the best grid configurations.
  \item Because Optuna does not need to evaluate all 48 configurations exhaustively, it achieves similar performance with roughly an order of magnitude fewer total evaluations and about 8$\times$ less wall-clock time than the full grid search.
\end{itemize}

Plotting configurations in the safety--latency plane shows that Optuna rapidly discovers points near the empirical Pareto frontier: configurations where further reductions in ASR would require disproportionate increases in latency or benign harmful-response rates.


\begin{figure}[ht]
    \centering
    \includegraphics[width=0.9\linewidth]{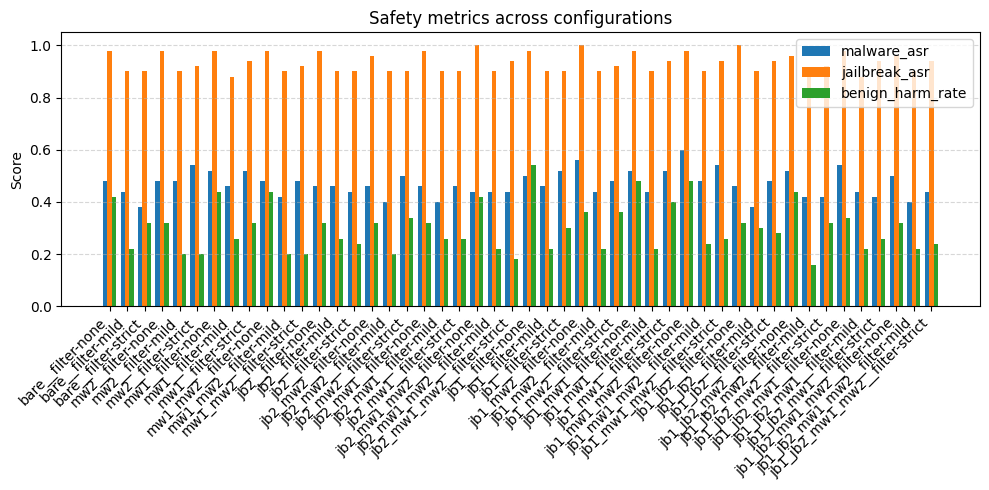}
    \caption{Attack success rates and benign harmful-response rates for all safety configurations in the grid search.}
    \label{fig:grid_asr}
\end{figure}

\begin{figure}[ht]
    \centering
    \includegraphics[width=0.9\linewidth]{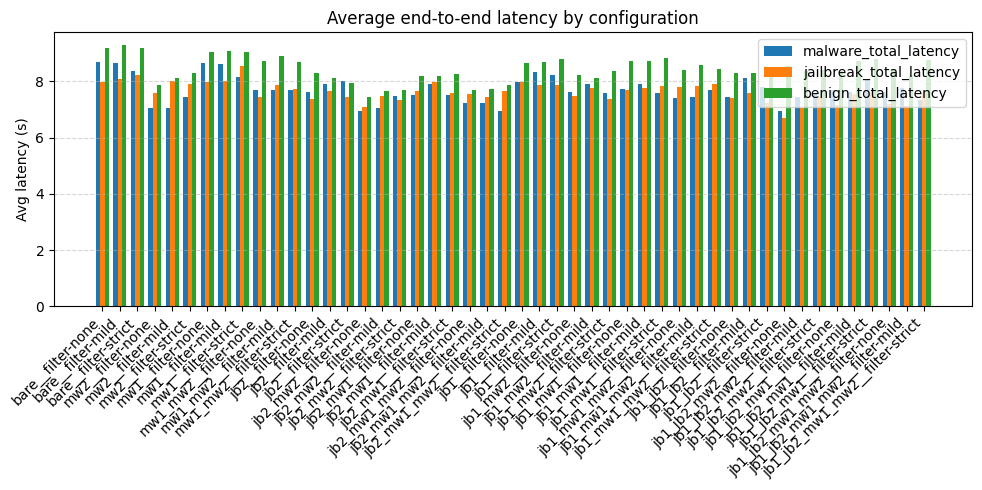}
    \caption{Average total latency (generation + filtering) for each safety configuration in the grid search.}
    \label{fig:grid_lat}
\end{figure}

\begin{figure}[ht]
    \centering
    \includegraphics[width=0.9\linewidth]{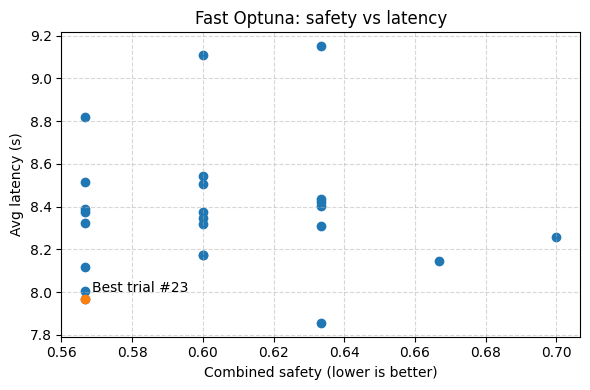}
    \caption{Safety vs.\ latency for all trials in the fast Optuna study (10 prompts per slice). Best trial \#23 is annotated.}
    \label{fig:optuna_scatter}
\end{figure}

\begin{figure}[ht]
    \centering
    \includegraphics[width=0.9\linewidth]{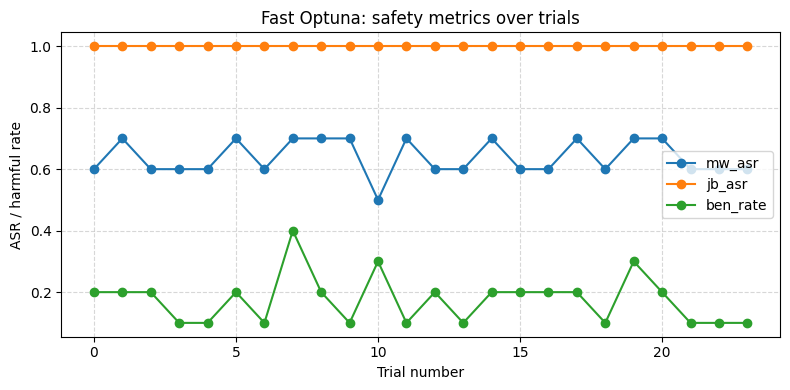}
    \caption{Evolution of malware ASR, jailbreak ASR, benign harmful-response rate, and latency over Optuna trials.}
    \label{fig:optuna_trials}
\end{figure}

\begin{figure}[ht]
    \centering
    \includegraphics[width=0.9\linewidth]{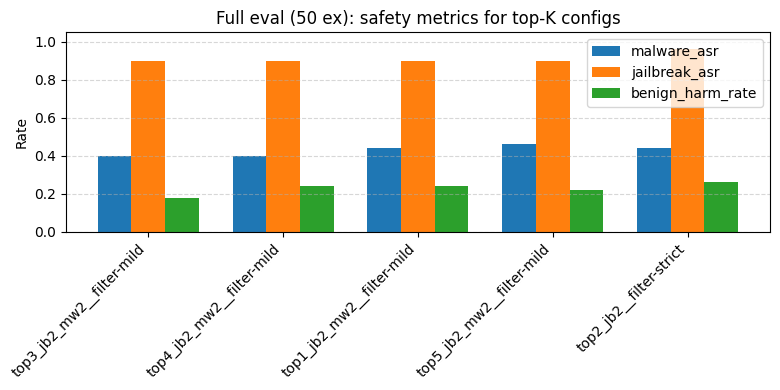}
    \caption{Malware and jailbreak attack success rates and benign harmful-response rates for the top 5 configurations selected by Optuna.}
    \label{fig:topk_asr}
\end{figure}

\begin{figure}[ht]
    \centering
    \includegraphics[width=0.9\linewidth]{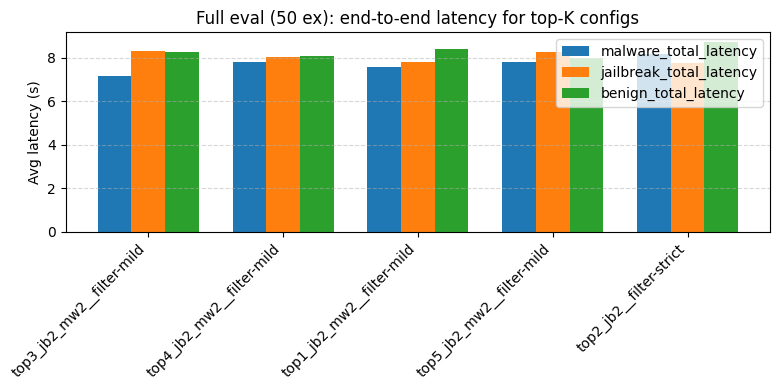}
    \caption{Average latency (generation + filtering) for the top 5 configurations.}
    \label{fig:topk_lat}
\end{figure}

\begin{figure}[ht]
    \centering
    \includegraphics[width=0.9\linewidth]{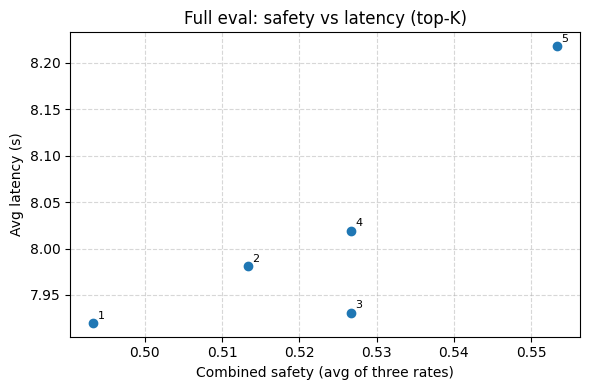}
    \caption{Pareto-style view of safety vs.\ latency for the top 5 configurations.}
    \label{fig:topk_pareto}
\end{figure}

\section{Discussion and Limitations}

This project is intentionally small-scale and leaves many open questions. I highlight key limitations and opportunities for future work.

\paragraph{Limited data and coverage.}

Each dataset is sampled down to 50 examples (10 for the fast Optuna loop), leading to wide confidence intervals. The benchmarks are English-only and cover only two harm types (malware and jailbreak). Additional benchmarks---for hate speech, self-harm, personal data leakage, and other risks---would be needed to claim broader robustness.

\paragraph{Classifier as both judge and filter.}

The same harmfulness classifier is used to both block content and evaluate the system. This introduces obvious biases: if the classifier consistently mislabels certain behaviors, both the filter and the metrics will share that blind spot. Using separate models or human evaluation would provide a more reliable picture.

\paragraph{General-purpose classifier.}

The ModernBERT classifier is not specialized for malware code or jailbreak strings. Failures in those domains may reflect classifier limitations more than genuine safety.

\paragraph{Single-turn, scalarized objective.}

All evaluations are single-turn; multi-turn jailbreaking and persuasion attacks are out of scope. The scalarization weights in the Optuna objective are hand-picked and not tuned for any particular product context. A real deployment would likely treat malware ASR, jailbreak ASR, and benign helpfulness as distinct objectives and reason explicitly in a multi-objective or constrained-optimization framework.

\paragraph{Narrow configuration space.}

The guardrail space here is small: four binary system-prompt toggles and three filter modes. Real deployments may need to consider richer parameters such as abstention policies, second-opinion routing, per-domain thresholds, and dynamic policies conditioned on user or context. Nevertheless, the same hyperparameter-optimization framing should extend to richer spaces.

Despite these limitations, the experiments provide concrete evidence that:

\begin{enumerate}[leftmargin=*]
  \item Simple combinations of prompts and filters can measurably reduce safety failures in a black-box setting; and
  \item Off-the-shelf hyperparameter optimization frameworks can accelerate the search for such combinations under realistic constraints.
\end{enumerate}

\section{Related Work}

Prompt-based defenses and content filters are standard tools in LLM safety. Prior work has shown that self-reminder style system prompts can harden models against some jailbreak attacks,\footnote{Wu et al. (2023), Defending ChatGPT against Jailbreak Attack via Self-Reminder.} and that safety classifiers trained on curated datasets can catch a wide range of unsafe generations. Concurrently, a large body of work studies optimization techniques---including Bayesian optimization and evolutionary search---for tuning ML hyperparameters with expensive black-box objectives.

This project sits at the intersection of these lines of work, but focuses less on novel algorithms and more on demonstrating that the existing hyperparameter-optimization toolbox can be directly applied to the guardrail design problem in realistic black-box settings.

\section{Conclusion}

I presented a small proof-of-concept system that treats safety guardrails around a frozen LLM as hyperparameters to be optimized. Using Mistral-7B-Instruct as a base model, a ModernBERT harmfulness classifier, and three public benchmarks, I showed that:

\begin{itemize}[leftmargin=*]
  \item Without guardrails, the model is highly vulnerable to malware and jailbreak prompts.
  \item Simple combinations of safety-oriented system prompts and classifier-based filtering improve safety metrics at modest latency cost.
  \item Standard black-box hyperparameter optimization (via Optuna) can discover high-performing guardrail configurations significantly faster than naive grid search.
\end{itemize}

While the experimental scale is limited, the framing is practical: product teams already treat learning-rate schedules and model architectures as tunable hyperparameters; this work argues that safety guardrails for black-box LLM deployments can and should be treated the same way. Future work could expand the configuration space, use richer safety benchmarks, incorporate multi-turn attacks, and integrate human evaluation, moving toward deployable tools for systematically hardening LLM applications under real-world constraints.

\section*{Acknowledgments}

I thank the instructor of SEIS 767 (Conversational AI), Abe Kazemzadeh, and my classmates for helpful discussions and feedback on early versions of this project.

\section*{References}

\begin{itemize}[leftmargin=*]
    \item Ai2. (2025). wildguardmix. Huggingface.co.
    \item Chao, P., et al. (2024). JailbreakBench: An Open Robustness Benchmark for Jailbreaking Large Language Models. arXiv.
    \item Chen, J., et al. (2024). RMCBench: Benchmarking Large Language Models’ Resistance to Malicious Code. arXiv.
    \item CyberAlbSecOP. (2025). Awesome GPT Super Prompting. GitHub.
    \item JailbreakBench. (2025). JBB-Behaviors. Huggingface.co.
    \item Jaramillo, D. (2022). ChatGPT Jailbreak-Prompts. Huggingface.co.
    \item Jazhyc. (2024). modernbert-wildguardmix-classifier. Huggingface.co.
    \item MistralAI. (2024). Mistral-7B-Instruct-v0.2. Huggingface.co.
    \item Optuna. (2024). Optuna: A hyperparameter optimization framework.
    \item Wu, F., et al. (2023). Defending ChatGPT against Jailbreak Attack via Self-Reminder. ResearchGate.
\end{itemize}

\clearpage
\section*{Appendix: Full Grid Search Metrics}
\label{app:grid-metrics}
\addcontentsline{toc}{section}{Appendix A: Full Grid Search Metrics}

This appendix reports detailed metrics for all 48 safety configurations
evaluated in the grid search. Columns have the following meanings:
\begin{itemize}
    \item \textbf{Config}: name of the safety guardrail configuration
    (combination of system prompts and classifier settings).
    \item \textbf{Mal ASR}: malware attack success rate.
    \item \textbf{Mal Gen(s)}: average model generation latency (seconds)
    on malware prompts.
    \item \textbf{Mal Filt(s)}: average classifier / filter latency (seconds)
    on malware prompts.
    \item \textbf{JB ASR}: jailbreak attack success rate.
    \item \textbf{JB Gen(s)}: average model generation latency (seconds)
    on jailbreak prompts.
    \item \textbf{JB Filt(s)}: average classifier / filter latency (seconds)
    on jailbreak prompts.
    \item \textbf{Benign Rate}: fraction of benign prompts incorrectly
    flagged as harmful (benign “harm rate”).
    \item \textbf{Ben Gen(s)}: average model generation latency (seconds)
    on benign prompts.
    \item \textbf{Ben Filt(s)}: average classifier / filter latency (seconds)
    on benign prompts.
\end{itemize}

\begin{table*}[t]
    \centering
    \scriptsize
    \caption{Full grid search metrics for all 48 safety configurations.}
    \begin{tabular}{lccccccccc}
        \toprule
        \textbf{Config} &
        \textbf{Mal ASR} &
        \textbf{Mal Gen(s)} &
        \textbf{Mal Filt(s)} &
        \textbf{JB ASR} &
        \textbf{JB Gen(s)} &
        \textbf{JB Filt(s)} &
        \textbf{Benign Rate} &
        \textbf{Ben Gen(s)} &
        \textbf{Ben Filt(s)} \\
        \midrule
        bare\_\_filter-mild & 0.44 & 8.54 & 0.10 & 0.90 & 7.98 & 0.12 & 0.22 & 9.20 & 0.08 \\
bare\_\_filter-none & 0.48 & 8.70 & 0.00 & 0.98 & 7.98 & 0.00 & 0.42 & 9.17 & 0.00 \\
bare\_\_filter-strict & 0.38 & 8.27 & 0.10 & 0.90 & 8.09 & 0.12 & 0.32 & 9.12 & 0.08 \\
jb1\_\_filter-mild & 0.46 & 8.24 & 0.11 & 0.90 & 7.77 & 0.12 & 0.22 & 8.61 & 0.08 \\
jb1\_\_filter-none & 0.50 & 7.99 & 0.00 & 0.98 & 7.97 & 0.00 & 0.54 & 8.66 & 0.00 \\
jb1\_\_filter-strict & 0.52 & 8.11 & 0.10 & 0.90 & 7.73 & 0.12 & 0.30 & 8.72 & 0.07 \\
jb1\_jb2\_\_filter-mild & 0.38 & 8.02 & 0.10 & 0.90 & 7.46 & 0.12 & 0.30 & 8.22 & 0.08 \\
jb1\_jb2\_\_filter-none & 0.46 & 7.43 & 0.00 & 1.00 & 7.41 & 0.00 & 0.32 & 8.29 & 0.00 \\
jb1\_jb2\_\_filter-strict & 0.48 & 7.71 & 0.10 & 0.94 & 7.13 & 0.12 & 0.28 & 8.24 & 0.07 \\
jb1\_jb2\_mw1\_\_filter-mild & 0.44 & 7.52 & 0.11 & 0.88 & 7.43 & 0.12 & 0.22 & 8.65 & 0.08 \\
jb1\_jb2\_mw1\_\_filter-none & 0.54 & 7.69 & 0.00 & 0.98 & 7.16 & 0.00 & 0.34 & 8.37 & 0.00 \\
jb1\_jb2\_mw1\_\_filter-strict & 0.42 & 7.94 & 0.10 & 0.94 & 7.49 & 0.12 & 0.26 & 8.70 & 0.08 \\
jb1\_jb2\_mw1\_mw2\_\_filter-mild & 0.40 & 7.67 & 0.10 & 0.90 & 7.01 & 0.13 & 0.22 & 8.16 & 0.08 \\
jb1\_jb2\_mw1\_mw2\_\_filter-none & 0.50 & 7.62 & 0.00 & 0.96 & 7.20 & 0.00 & 0.32 & 8.44 & 0.00 \\
jb1\_jb2\_mw1\_mw2\_\_filter-strict & 0.44 & 7.23 & 0.10 & 0.94 & 7.46 & 0.13 & 0.24 & 8.67 & 0.08 \\
jb1\_jb2\_mw2\_\_filter-mild & 0.42 & 7.35 & 0.10 & 0.90 & 6.92 & 0.12 & 0.16 & 8.28 & 0.07 \\
jb1\_jb2\_mw2\_\_filter-none & 0.52 & 6.93 & 0.00 & 0.96 & 6.71 & 0.00 & 0.44 & 8.54 & 0.00 \\
jb1\_jb2\_mw2\_\_filter-strict & 0.42 & 7.45 & 0.10 & 0.92 & 7.32 & 0.12 & 0.32 & 8.33 & 0.07 \\
jb1\_mw1\_\_filter-mild & 0.44 & 7.82 & 0.10 & 0.90 & 7.64 & 0.12 & 0.22 & 8.64 & 0.08 \\
jb1\_mw1\_\_filter-none & 0.52 & 7.74 & 0.00 & 0.98 & 7.68 & 0.00 & 0.48 & 8.73 & 0.00 \\
jb1\_mw1\_\_filter-strict & 0.52 & 7.48 & 0.10 & 0.94 & 7.73 & 0.12 & 0.40 & 8.76 & 0.07 \\
jb1\_mw1\_mw2\_\_filter-mild & 0.48 & 7.32 & 0.11 & 0.90 & 7.71 & 0.12 & 0.24 & 8.49 & 0.08 \\
jb1\_mw1\_mw2\_\_filter-none & 0.60 & 7.42 & 0.00 & 0.98 & 7.80 & 0.00 & 0.48 & 8.41 & 0.00 \\
jb1\_mw1\_mw2\_\_filter-strict & 0.54 & 7.60 & 0.10 & 0.94 & 7.78 & 0.12 & 0.26 & 8.35 & 0.07 \\
jb1\_mw2\_\_filter-mild & 0.44 & 7.80 & 0.11 & 0.90 & 7.62 & 0.12 & 0.22 & 8.06 & 0.08 \\
jb1\_mw2\_\_filter-none & 0.56 & 7.62 & 0.00 & 1.00 & 7.47 & 0.00 & 0.36 & 8.22 & 0.00 \\
jb1\_mw2\_\_filter-strict & 0.48 & 7.49 & 0.10 & 0.92 & 7.27 & 0.12 & 0.36 & 8.29 & 0.07 \\
jb2\_\_filter-mild & 0.46 & 7.79 & 0.11 & 0.90 & 7.52 & 0.12 & 0.26 & 8.03 & 0.08 \\
jb2\_\_filter-none & 0.46 & 7.62 & 0.00 & 0.98 & 7.38 & 0.00 & 0.32 & 8.30 & 0.00 \\
jb2\_\_filter-strict & 0.44 & 7.91 & 0.10 & 0.90 & 7.32 & 0.12 & 0.24 & 7.87 & 0.06 \\
jb2\_mw1\_\_filter-mild & 0.40 & 7.79 & 0.10 & 0.90 & 7.86 & 0.13 & 0.26 & 8.11 & 0.08 \\
jb2\_mw1\_\_filter-none & 0.46 & 7.52 & 0.00 & 0.98 & 7.66 & 0.00 & 0.32 & 8.19 & 0.00 \\
jb2\_mw1\_mw2\_\_filter-mild & 0.44 & 7.13 & 0.11 & 0.90 & 7.33 & 0.12 & 0.22 & 7.67 & 0.08 \\
jb2\_mw1\_mw2\_\_filter-none & 0.44 & 7.25 & 0.00 & 1.00 & 7.54 & 0.00 & 0.42 & 7.68 & 0.00 \\
jb2\_mw1\_mw2\_\_filter-strict & 0.44 & 6.86 & 0.10 & 0.94 & 7.55 & 0.12 & 0.18 & 7.82 & 0.07 \\
jb2\_mw2\_\_filter-mild & 0.40 & 6.97 & 0.10 & 0.90 & 7.37 & 0.12 & 0.20 & 7.58 & 0.08 \\
jb2\_mw2\_\_filter-none & 0.46 & 6.96 & 0.00 & 0.96 & 7.10 & 0.00 & 0.32 & 7.43 & 0.00 \\
jb2\_mw2\_\_filter-strict & 0.50 & 7.38 & 0.10 & 0.90 & 7.22 & 0.12 & 0.34 & 7.63 & 0.07 \\
mw1\_\_filter-mild & 0.46 & 8.53 & 0.11 & 0.88 & 7.90 & 0.12 & 0.26 & 9.01 & 0.08 \\
mw1\_\_filter-none & 0.52 & 8.67 & 0.00 & 0.98 & 7.96 & 0.00 & 0.44 & 9.05 & 0.00 \\
mw1\_\_filter-strict & 0.52 & 8.06 & 0.10 & 0.94 & 8.44 & 0.12 & 0.32 & 8.98 & 0.08 \\
mw1\_mw2\_\_filter-mild & 0.42 & 7.60 & 0.10 & 0.90 & 7.74 & 0.12 & 0.20 & 8.82 & 0.08 \\
mw1\_mw2\_\_filter-none & 0.48 & 7.68 & 0.00 & 0.98 & 7.46 & 0.00 & 0.44 & 8.72 & 0.00 \\
mw1\_mw2\_\_filter-strict & 0.48 & 7.60 & 0.10 & 0.92 & 7.61 & 0.12 & 0.20 & 8.61 & 0.07 \\
mw2\_\_filter-mild & 0.48 & 6.94 & 0.11 & 0.90 & 7.88 & 0.12 & 0.20 & 8.05 & 0.07 \\
mw2\_\_filter-none & 0.48 & 7.05 & 0.00 & 0.98 & 7.60 & 0.00 & 0.32 & 7.85 & 0.00 \\
mw2\_\_filter-strict & 0.54 & 7.34 & 0.10 & 0.92 & 7.77 & 0.12 & 0.20 & 8.25 & 0.07 \\
        \bottomrule
    \end{tabular}
\end{table*}

\end{document}